\documentclass{elsart}
\usepackage{amssymb}
\usepackage{latexsym}
\usepackage{amssymb, amsmath,hyperref}
\newcommand{\xx}{U}
\newtheorem{nprop}{Proposition}[{section}]
\newtheorem{nrem}[nprop]{Remark}
\newtheorem{nexmp}[nprop]{Example}
\newtheorem{nthm}[nprop]{Theorem}
\newtheorem{ncor}[nprop]{Corollary}
\newtheorem{ndefn}[nprop]{Definition}
\begin{document}

\begin{frontmatter}

  \title{$R$-separation of variables for the conformally invariant Laplace equation}
  \author[A]{Mark \ Chanachowicz},
  \ead{mchanach@math.uwaterloo.ca}
  \author[B]{Claudia M.\ Chanu\corauthref{cor}}
  \corauth[cor]{Corresponding author.}
  \ead{claudiamaria.chanu@unito.it}
  \author[A]{Raymond G.\ McLenaghan}
  \ead{rgmclena@uwaterloo.ca}
  \address[A]{Department of Applied Mathematics, University of Waterloo,
    Waterloo, Ontario N2L~3G1, Canada}
  \address[B]{Dipartimento di Matematica, Universit$\grave{a}$ di Torino, 
    via Carlo Alberto 10, 10123 Torino, Italy}
%\author{M. CHANACHOWICZ, R. G. McLENAGHAN}
%
%\address{University of Waterloo, Waterloo, Ontario, Canada, Inc, N2L 3G1\\
%E-mail: mchanach@math.uwaterloo.ca, rgmclena@uwaterloo.ca}
%
%\author{C. CHANU}
%
%\address{Dipartimento di Matematica, Universit\grave{a} di Torino, 
%    via Carlo Alberto 10, 10123 Torino, Italy} 
%E-mail: claudiamaria.chanu@unito.it}  

\begin{abstract}
The conditions for $R$-separation of variables for the conformally invariant Laplace equation on
an $n$-dimensional Riemannian manifold are determined  and compared with the conditions for the additive separation of the null geodesic Hamilton-Jacobi equation.  The case of 3-dimensions is examined in
detail and it is proven that on any conformally flat manifold the two equations separate in the same coordinates.
\end{abstract}

\begin{keyword}
$R$-separation \sep conformal separation \sep Laplace equation \sep conformal flatness \sep St\"{a}ckel separability

%Go to www.aip.org/pacs to find the code.
\PACS 02.30.Jr \sep 02.40.Ky 
%(PDE) (Riemannian geometry)
\end{keyword}
\end{frontmatter}

\section{Introduction}\label{intro}
The subject of this paper is the study of $R$-separation of variables for the conformally
invariant Laplace equation on an $n$-dimensional Riemannian manifold $(M,\mathbf g)$.  
This equation may be written as
\begin{equation}\label{CILeq}
\mathbb{H}\psi:=\Delta \psi +\frac 14\left(\frac{n-2}{n-1}\right)R_s\psi = 0,
\end{equation}
where
\begin{equation}\label{LBop}
\Delta \psi  := \sqrt{g}^{-1}\partial_i(\sqrt{g}g^{ij}\partial_j\psi)
\end{equation}
is the Laplace-Beltrami operator and $R_s$ is the Ricci scalar.
A closely related problem is the study of additive separation of variables for the Hamilton-Jacobi
equation for the null geodesics (in the pseudo-Riemannian case) namely
\begin{equation}\label{NGeq}
g^{ij}\partial_{i}W\partial_{j}W=0.
\end{equation}
An important property of both (\ref{CILeq}) and (\ref{NGeq}) is invariance under conformal transformation
of the metric
\begin{equation}\label{CTmetric}
\tilde{\mathbf g}=e^{2\sigma}{\mathbf g}.
\end{equation}
It follows that if $\psi$ is any solution of $\mathbb{H}\psi=0$, then $\tilde{\psi}=e^{\frac{2-n}{2}}\psi$
is a solution of $\tilde{\mathbb{H}}\tilde{\psi}=0$ on any conformally related manifold.  Consequently,
$R$-separability of the CI-Laplace equation is a conformally invariant property.  This property is
not shared by the Laplace equation
\begin{equation}\label{Leq}
\Delta{\psi}=0,
\end{equation}
which is the equation most often studied in this regard \cite{Bocher,KeM,Rsep}. Note that  (\ref{Leq})
is the equation usually considered as the extension of the ordinary Laplace equation,
\begin{equation}\label{OLeq}
\dfrac{\partial^2 \psi}{\partial x_1^2} + \ldots + \dfrac{\partial^2 \psi}{\partial x_n^2}=0,
\end{equation}
to a general non-flat Riemannian manifold $(M, \mathbf{g})$.   However, we remark that  both (\ref{CILeq}) and (\ref{Leq})
reduce to the ordinary Laplace equation on any flat manifold.

In this paper we shall demonstrate the advantages of studying $R$-separability for the CI-Laplace
equation continuing the work begun in \cite{Ray}.  In Section \ref{sec_2}, based on the results of \cite{Rsep,HJE},
we derive necessary and sufficient conditions for $R$-separation in the coordinates $(q^{i})$ in terms
of the components of the contravariant metric tensor $(g^{ij})$.  The precise result is given in
Theorem \ref{t_Rsep}.  In Section \ref{sec_3} we consider the case $n=3$. We first employ conformal invariance
and St\"{a}ckel theory to derive the form of the contravariant metric in general conformally
separable coordinates.  We call a coordinate system \textit{general} if it contains no conformally ignorable 
coordinate.  We next obtain  the contravariant metric in general $R$-separable coordinates in terms of an
arbitrary fifth degree polynomial function of the coordinates $q^{i}$ by imposing the compatibility
condition of Theorem \ref{t_Rsep}.  It follows that $R$-separation of (\ref{CILeq})
occurs in general conformally separable coordinates if and only if the metric is conformally flat.
We conclude this section by obtaining conditions for $R$-separation on the Laplace equation (\ref{Leq})
on conformally flat spaces. In Section \ref{sec_4} we show that on the locally conformally flat 3-sphere $\mathbb S_3$
the Laplace equation (\ref{Leq}) admits $R$-separation only if the coordinates are separable, while for
the CI-Laplace equation there exist proper conformally separable coordinates allowing fixed energy
$R$-separation.  This example provides a justification for studying $R$-separation for the conformally
invariant equation.

For the flat
 case (in which the CI-Laplacian becomes the classical one) we recover the
 results  given by B\^{o}cher \cite{Bocher} and Boyer et al. \cite{KeM} about the conformal factor such that the metric in general $R$-separable coordinates is flat and the transformation to Cartesian
coordinates.
Furthermore, we applied these results to provide CI-Laplace $R$-separable
coordinates on other conformally flat manifolds.
Section \ref{sec_5} contains the conclusion.   

\section{The CI-Laplace equation and $R$-separation}\label{sec_2} 
We define $R$-separation for  the CI-Laplace equation, according to \cite{Rsep}, as follows:

\begin{ndefn}\rm
Multiplicative $R$-separation of a single second order partial differential equation (PDE) is the search for a solution $\psi$ of the form 
\begin{equation}\label{Rdef}
\psi=R(q^1,\ldots, q^n)\prod_i \phi_i(q^i,c_a) \qquad c_a\in \mathbb R \quad a=1,\ldots,{2n-1};
\end{equation}
satisfying the completeness condition
  $$
\mathrm{rank}\left[ \frac{\partial}{\partial c_a}\!\left(\frac{\phi_i^\prime}{\phi}\right)\ \frac{\partial}{\partial c_a}\!\left(\frac{\phi_i^{\prime\prime}}{\phi}\right)\right]=2n-1, 
\qquad 
a=1,\ldots, 2n-1, \quad i=1,\ldots, n.
$$ 
\end{ndefn}

\begin{nrem} \rm
The completeness condition is equivalent to the fact  that for any choice of the $2n-1$ values $c_a$ there is a unique $R$-separated
solution  such that at a point $q_0\in M$ 
$$\left(\frac{\phi_i^\prime}{\phi},\; \frac{\phi_j^{\prime\prime}}{\phi}\right)_{q_0}=(c_a)\qquad (i=1,\ldots, n, \ j=1,\ldots, n-1). $$
Moreover, the PDE splits into $n$ separated ordinary differential equations (see \cite{Moon_art,Moon}).
The constants $(c_a)$ are of two types: $n-1$ of them are separating constants, involved into the separated ordinary differential equations (ODE)s, the other $n$ are integration constants, arising from the integration of the ODEs (see \cite{Rsep}). 
\end{nrem}

The study of $R$-separation of the CI-Laplace equation, instead of the classical equation, seems more natural
\cite{Ray}. Indeed, the following property holds.

\begin{nprop}
The existence of a complete $R$-separated solution of the CI-Laplace equation is a conformally invariant property that holds on the whole class of conformally related metrics.
\end{nprop}
\begin{pf}
Follows directly from (\ref{Rdef}) and the conformal invariance of (\ref{CILeq}).
\qed
\end{pf}
In order to obtain necessary and sufficient conditions for the existence of a complete $R$-separated solution
in a given coordinate system, 
first we transform $R$-separation into the multiplicative separation of a related PDE
involving the function $R$: 
$R$-separated solutions of $\mathbb{H}\psi=0$ correspond to
multiplicatively separated solutions $$\phi=\frac{\psi}R=\textstyle\prod_i\phi_i(q^i,c_a)$$ of
$$
\Delta \phi+ 2\nabla \ln R \cdot \nabla \phi +\xx \phi=0, 
$$
where $\xx$ is the modified potential
\begin{equation}
\xx:=\frac 14\left(\frac{n-2}{n-1}\right)R_s +\frac {\Delta R}{R}.
\end{equation}
By applying the techniques giving differential conditions for the $R$-separation of a single PDE \cite{Rsep}
we arrive at:

\begin{nthm} \label{t_Rsep}
Equation $(\ref{CILeq})$ 
admits $R$-separation in the coordinates $(q^i)$ if and only if 
\begin{enumerate}
\item the coordinates are orthogonal: $g_{ij} = 0, \ \ i \neq j$; 
\item the contravariant components $(g^{ii})$ satisfy the differential condition  
\begin{equation}\label{pino}
\frac{S_{ij}(g^{hh})}{g^{hh}}=\frac{S_{ij}(g^{kk})}{g^{kk}}, \qquad (\forall\ h,k,\ \forall\; i\neq j, \  i,j \; \mathrm{n.s.})
\end{equation}
where $S_{ij}$ are second order operators, called {\em St\"ackel operators}, defined as
\begin{equation}
S_{ij}(f)= \partial_i\partial_j f-\partial_i\ln |g^{jj}|\partial_j f -\partial_j\ln |g^{ii}|\partial_i f;
\end{equation} 
\item \label{Rcon}
the function $R$ is (up to separated factors) a solution of 
\begin{equation}\label{sys}
	\partial_i \ln R=\tfrac 12 \Gamma_i, 
\end{equation}
where $\Gamma_i=g^{hk}\Gamma_{hki}$;
\item \label{Xicon} 
the modified potential $\xx$ is a pseudo-St\"ackel factor i.e., it
is of the form
$\xx=g^{hh}f_h(q^h)$ for suitable functions $f_h$.
\end{enumerate}
\end{nthm}
\begin{pf}
We apply the conditions for $R$-separation of the fixed energy $R$-separation of the Schr\"odinger equation
$$
-\frac{\hbar^2}2 \Delta \psi + (V-E)\psi=0
$$
given in \cite{Rsep} to equation (\ref{CILeq}), that is for $E=0$ and 
$V=-\dfrac {\hbar^2}{8}\left(\dfrac{n-2}{n-1}\right)R_s$.
\qed
\end{pf}

\begin{nrem}\label{r6} \rm
Orthogonal coordinates satisfying condition (\ref{pino}) are called {\em conformally separable} (see \cite{HJE}), while orthogonal coordinates satisfying 
$S_{ij}(g^{hh})=0$ are said to be {\em simply separable}. 
The additive separation of variables for the null geodesic Hamilton-Jacobi equation in orthogonal coordinates,  
\begin{equation}\label{ONgHJ}
g^{ii}(\partial_iW)^2=0, 
\end{equation}
and for the geodesic Hamilton-Jacobi equation
\begin{equation}\label{OgHJ}
\tfrac 12 g^{ii}(\partial_iW)^2=E, \quad (E\in \mathbb R), 
\end{equation}
occurs if and only if the coordinates are conformally separable and simply separable, respectively. This fact shows an important link between Eq. (\ref{CILeq}) and Eq. (\ref{NGeq}).
\end{nrem}	
Also in the Riemannian case, even if the null geodesics are
trivial, the study of conformal separation can be applied effectively to the
CI-Laplace equation.
Indeed, as for the Laplace equation $\Delta \psi=0$, we have that

\begin{ncor}
A necessary condition for $R$-separation of the CI-Laplace equation $(\ref{CILeq})$ in a given coordinate system is that the null geodesic equation $(\ref{NGeq})$ is additively separable in the same coordinates.
\end{ncor}  

\begin{nrem} \rm
Two conditions equivalent to (\ref{pino}) are
\begin{itemize}
\item $\mathbf g$ is conformal to a metric which is separable for the geodesic Hamilton-Jacobi equation
(\ref{OgHJ})
 in the same coordinates; 
\item there exists a St\"ackel matrix $S$ (that is a regular $n\times n$ matrix of which the elements of the $i$th-row depend only on $q^i$) such 
that \cite{Moon_art} 
\begin{equation}\label{min}
\frac{g^{ii}}{g^{jj}}=\frac{M_{in}}{M_{jn}},
\end{equation}
where $M_{in}$ is the minor of $S$ obtained by eliminating the $i$-th row and the $n$-th column.
We remark that the elements of the last column of the St\"ackel matrix are not involved in (\ref{min}).
\end{itemize}
\end{nrem}

\begin{nrem} \rm
The PDE system  (\ref{sys}) determines the form of the factor $R$; the integrability conditions of the system are satisfied when the metric is conformally separable. 
\end{nrem}

\begin{nrem}\rm
By inserting the form of $R$ in the modified potential $\xx$ we get
$$\xx=\dfrac 14 g^{ii}(2\partial_i \Gamma_i-\Gamma_i^2)+\dfrac 14\left(\dfrac{n-2}{n-1}\right)R_s.$$
Two equivalent forms of the compatibility condition (\ref{Xicon}) are the following:
\begin{itemize}
\item
the conformal metric $\xx^{-1}{g^{hh}}$ is a separable metric;
\item condition
${S_{ij}(g^{hh})}\xx={S_{ij}(\xx)}{g^{hh}}$ holds.
\end{itemize} 
\end{nrem} 

The conditions are formally the same as for the classical Laplace equation, except for the presence of the Ricci scalar in the modified potential.
The condition that the coordinates are conformally separable holds for all conformally related metrics.
Moreover, because of the term containing $R_s$ in $\xx$, the compatibility
condition is also satisfied (or not) on the whole class of conformally related metrics. 

\begin{nprop} Let $\tilde \xx$ be the modified potential associated with $\tilde g^{hh}={e^{-2\sigma}}{g^{hh}}$.
Then $\tilde \xx={e^{-2\sigma}} {\xx}$ and it is a pseudo-St\"ackel factor if and only if $\xx$ is.
\end{nprop}

\section{The three-dimensional case}\label{sec_3}
The CI-Laplace equation on a three dimensional manifold is 
\begin{equation}\label{CIL3d}
\Delta \psi +\frac 18 R_s \psi=0.
\end{equation}

\begin{ndefn} \rm
A coordinate $q^i$ is {\em conformally ignorable} if it appears in the conformal factor of the metric only, that is if $\partial_i(g^{hh}/g^{kk})=0$ for all $h,k$. We call a coordinate system {\em general} if it does not contain any conformally ignorable coordinate. 
\end{ndefn}

Up to a coordinate transformation of the form $\tilde q^i(q^i)$,
a coordinate $q^i$ is conformally ignorable if and only if 
$\partial_i$ is a conformal Killing vector, that is an infinitesimal conformal symmetry.
We restrict ourselves to general coordinate systems, leaving as a further development the analysis of the
cases involving conformal symmetries.  

The form of the general conformally separable coordinates in a three dimensional manifold is given in the
following proposition (see \cite{KeM}) 
\begin{nprop} \label{p_csep}
In general conformally separable coordinates $(q^i)$, the form of the (contravariant) metric on a 3-manifold is given
by
\begin{equation} \label{met_csep}
g^{ii}=Q h_i(q^i)(q^{i+2}-q^{i+1}) \qquad {\mbox{\small $i=1,\ldots,3\ (mod \ 3)$}}
\end{equation}
where $Q$ is the conformal factor, and $h_i$ three arbitrary functions of a single variable.
\end{nprop}

\begin{pf}
Following  \cite{KeM},
without loss of generality, we may choose $S$ to be a $3\times 3$ St\"ackel matrix with third column set equal to unity:
\begin{equation}
S=\left[\begin{matrix}
\phi_1\ \ & \psi_1\ \ & 1  \\
\phi_2  & \psi_2 & 1 \\
\phi_3  & \psi_3 & 1 \\
\end{matrix}
\right]
\end{equation}
Then,  we have
\begin{equation}
g^{11}=Q(\psi_3\phi_2-\psi_2\phi_3),\quad  g^{22}=Q(\psi_1\phi_3-\psi_3\phi_1),\quad
g^{33}=Q(\psi_2\phi_1-\psi_1\phi_2).
\end{equation}
In the general case, we can assume that none of the $\psi_i$ and $\phi_i$ are identically null. Thus
\begin{equation}
g^{11}\!=Q\phi_2\phi_3\!\left(\frac{\psi_3}{\phi_3}-\frac{\psi_2}{\phi_2}\right)\!, \;
g^{22}\!=Q\phi_1\phi_3\!\left(\frac{\psi_1}{\phi_1}-\frac{\psi_3}{\phi_3}\right)\!, \;
g^{33}\!=Q\phi_1\phi_2\!\left(\frac{\psi_2}{\phi_2}-\frac{\psi_1}{\phi_1}\right)\!.
\end{equation}
By transforming each coordinate $\tilde q^i=\tilde q^i(q^i)$  and the conformal factor such that  
$$\tilde g^{ii} \to \phi_i g^{ii} \qquad \tilde Q \to Q \phi_1\phi_2\phi_3,$$
we get $g^{ii}=Q(F_{i+2}-F_{i+1})$ with $F_i(q^i)=\dfrac{\psi_i}{\phi_i}$.
If none of the $F_i$ is a constant, then we can use them as coordinates; thus, we obtain
$$
g^{11}=Qh_1(q^1)(q^3-q^2), \ \ \ g^{22}=Qh_2(q^2)(q^1-q^3), \ \ \  g^{33}=Qh_3(q^3)(q^2-q^1). \qed
$$
\end{pf}

\begin{nrem}\rm
If one of the elements of the St\"ackel matrix is zero or one of the functions $F_i$ is a constant, then,
up to a coordinate transformation $\tilde q^i(q^i)$, one of the coordinates is {conformally ignorable}.
\end{nrem}
By Proposition \ref{p_csep} and Theorem \ref{t_Rsep} we arrive at
\begin{nthm}\label{t_rsep}
The form of the metric in general $R$-separable coordinates for the CI-Laplace equation is
\begin{equation}\label{m_Rsep}
g^{ii}=Q P(q^i)\cdot(q^{i+2}-q^{i+1}) \qquad {\mbox{\small $i=1,\ldots,3\ (mod \ 3)$}}
\end{equation}
where $P$ is an arbitrary fifth-degree polynomial.
\end{nthm}
\begin{pf}
Computing the modified potential $\xx$ for the 
general conformal separable 
metric (\ref{met_csep})
and
imposing the compatibility condition $$S_{ij}(g^{hh})\xx=S_{ij}(\xx)g^{hh},$$ we get three additional differential conditions on the functions
$h_i$ that form a linear second order ODE system in three unknowns, whose solution is $h_i=P(q^i)$, where $P$
is an arbitrary fifth-degree polynomial.
\qed
\end{pf}

It is interesting to note that the compatibility condition has an intriguing geometrical interpretation.

\begin{nthm}
On a 3-manifold, $R$-separation of the CI-Laplace equation occurs in general conformal separable coordinates
if and only if the metric is conformally flat.
\end{nthm}

\begin{pf}
The conformal flatness conditions for a 3-dimensional Riemannian manifold are (see for instance \cite{Eisen_book})
\begin{equation}
R_{ijk}= \nabla_k R_{ij} -\nabla_j R_{ik} +\tfrac 14(g_{ik}\nabla_jR_s-g_{ij}\nabla_kR_s) =0, 
\end{equation}
where $\nabla$ is the covariant derivative and $R_{ij}$ the covariant Ricci tensor.
By imposing these conditions on the general conformal separable metric (\ref{met_csep}), we get three independent second order linear ODEs in the $h_i$ which are equivalent to those allowing $R$-separation. 
Thus, the conformal flatness condition is equivalent to the compatibility condition for $R$-separation.
\qed
\end{pf}

\begin{nrem} \rm
If one or more conformally ignorable coordinates appear, then being conformally flat is a sufficient
but no longer a necessary condition for $R$-separation. Hence, in particular, equations (\ref{CILeq})
and (\ref{NGeq}) separate in the same orthogonal coordinates for all conformally flat 3-manifolds.
\end{nrem}

We can apply these results to the study of $R$-separation  for the classical Laplace equation:

\begin{nthm}
If a three dimensional manifold satisfies $R_s=0$, then $R$-separation of the Laplace equation $\Delta\psi=0$ occurs in general coordinates 
if and only if the manifold is conformally flat.
\end{nthm}
\begin{pf}
Since $R_s=0$, the CI-Laplace equation and Laplace equation coincide. \qed
\end{pf}

\begin{nthm}
On a conformally flat three dimensional manifold, $R$-separa\-tion of the Laplace equation
$\Delta\psi=0$ occurs in general conformally separable coordinates 
if and only if the Ricci scalar $R_s$ satisfies the compatibility condition
$$
S_{ij}(g^{hh})R_s=S_{ij}(R_s)g^{hh}.
$$
\end{nthm}
\begin{pf}
Since the manifold is conformally flat, the CI-Laplace equation admits $R$-separation of variables in 
general conformally separable coordinates, that is the modified potential 
$$\xx=\frac{\Delta R}R +\frac{R_s}8$$
satisfies ${S_{ij}(g^{hh})}\xx={S_{ij}(\xx)}{g^{hh}}$. Then, the modified potential for the Laplace equation
$\xx_L=\frac{\Delta R}R$ satisfies the same condition if and only if $R_s$ does. \qed
\end{pf}

\section{Applications and examples}\label{sec_4}
In this section we apply the discussed techniques to examples on the sphere and other conformally flat three-dimensional manifolds.
\begin{nexmp}\rm
The 3-sphere $\mathbb S_3$ is a conformally flat 3-manifold with constant Ricci scalar $R_s>0$.
Hence, $R$-separation of the Laplace equation occurs only in those conformally separable coordinates
such that 
$$
S_{ij}(g^{hh})R_s=S_{ij}(R_s)g^{hh}=0,
$$
since $R_s$ is a constant.
Thus, the metric components must satisfy $S_{ij}(g^{hh})=0$ and the coordinates are necessarily
separable (see Remark \ref{r6}). Due to this additional condition on the metric (\ref{m_Rsep}) the terms of degree five and four of the polynomial
$P$ must vanish and $P$ reduces to a third degree polynomial. 
Moreover, by examining the conditions for the $R$-separation of the Helmholtz equation (see \cite{KeMTo})
$$\Delta\psi=E\psi \qquad (E\in \mathbb R),$$
we have that,
on $\mathbb S_3$, the Laplace equation, $$\Delta\psi=0,$$ admits $R$-separation only in the general separable coordinates such that $R$-separation of the
Helmholtz equation occur.

Finally, we remark that on $\mathbb S_3$ the CI-Laplace equation (\ref{CIL3d})
coincides with the Helmholtz equation for the fixed value of the energy 
$$
E=-\tfrac 18 R_s.
$$
Hence, locally there exist proper conformally separable coordinates allowing 
fixed energy $R$-separation of the Helmholtz equation for the value $E=-R_s/8$. For details see Example \ref{example_20}.
\end{nexmp}

A fundamental example is the flat case, where the Laplace equation and the CI-Laplace equation are the same; it has been treated by several authors (see \cite{Bocher,MorseFeshb,Moon_art,KeM}).

\begin{nexmp}\rm
On the Euclidean three dimensional space $\mathbb E^3$ the CI-Laplace equation is $\Delta \psi=0$.
In order to determine the expression of general $R$-separable coordinates on $\mathbb E^3$ we need to compute
the conformal factor $Q$ such that the metric (\ref{m_Rsep}) is flat and the coordinate transformations from
a Cartesian coordinate system. 
We restrict ourselves to the case where all roots of $P$ are real and distinct (see \cite{Bocher} for the detailed analysis of all the possibilities).
Let us denote by $(q^1,q^2,q^3)$ the $R$-separable coordinates,  $(x^1,x^2,x^3)$ the Cartesian coordinates 
 and by $e_1<e_2<e_3<e_4<e_5$ the five roots of the polynomial $P$. By using pentaspherical coordinates (see 
 \cite{Bocher,MorseFeshb,KeM}) we can derive
the following relations linking Cartesian coordinates to the $R$-separable ones (we adopt the same notation as
in \cite{KeM})
\begin{eqnarray}\label{Kalnins_formula_b}
\lambda\cdot{x^1} &=& \sqrt{\frac{(q^1 - e_2)\cdot(q^2 - e_2)\cdot(q^3 - e_2)}{(e_2 - e_1)\cdot(e_2 - e_3)\cdot(e_2 - e_4)\cdot(e_2 - e_5)}}\nonumber\\
\lambda\cdot{x^2} &=& \sqrt{\frac{(q^1 - e_3)\cdot(q^2 - e_3)\cdot(q^3 - e_3)}{(e_3 - e_1)\cdot(e_3 - e_2)\cdot(e_3 - e_4)\cdot(e_3 - e_5)}}\\
\lambda\cdot{x^3} &=& \sqrt{\frac{(q^1 - e_4)\cdot(q^2 - e_4)\cdot(q^3 - e_4)}{(e_4 - e_1)\cdot(e_4 - e_2)\cdot(e_4 - e_3)\cdot(e_4 - e_5)}}\nonumber
\end{eqnarray}
where
\begin{eqnarray}\label{Kalnins_formula_a}
\lambda &=& \sqrt{\frac{(q^1 - e_1)\cdot(q^2 - e_1)\cdot(q^3 - e_1)}{(e_1 - e_2)\cdot(e_1 - e_3)\cdot(e_1 - e_4)\cdot(e_1 - e_5)}}\nonumber\\ &+& \sqrt{\frac{-(q^1 - e_5)\cdot(q^2 - e_5)\cdot(q^3 - e_5)}{(e_5 - e_1)\cdot(e_5 - e_2)\cdot(e_5 - e_3)\cdot(e_5 - e_4)}}
\end{eqnarray}
and 
$$
e_1<q^1<e_2<q^2<e_3<q^3<e_4<e_5.
$$
It can be checked directly by expressing the Euclidean metric $\mathbf{g}_E$ with respect to the coordinates $(q^i)$
that we obtain
\begin{eqnarray*}
\mathbf{g}_E=\delta_{ij} dx^i\otimes dx^j &=&
\delta_{ij}\frac{-(q^i-q^{i+1})(q^i-q^{i+2})}{4 P(q^i) \lambda^2} dq^i\otimes dq^j \\
&=&\delta_{ij}\frac{\prod_h(q^h-q^{h+1})}{4 P(q^i) \lambda^2 (q^{i+2}-q^{i+1})} dq^i\otimes dq^j,
\end{eqnarray*}
which is of the form (\ref{m_Rsep}) with conformal factor $Q_E$ given by
\begin{equation}
Q_E= \frac{4\lambda^2}{\prod_h(q^h-q^{h+1})},
\end{equation}
according to the formulas given in \cite{Bocher} and  in \cite{KeM} (where a factor of 4 seems to be missing).
\end{nexmp}

\begin{nrem} \rm
The conformal metric 
$$
\tilde{g}_{ii}=\frac{(q^i-q^{i+1})(q^i-q^{i+2})}{P(q^i)} 
$$
is the general three-dimensional conformally flat metric allowing multiplicative separation of the Helmholtz equation
computed by Eisenhart \cite{Eisen35}. 
\end{nrem}

The formulas for the flat case can be adapted to a general conformally flat manifold $(M, \mathbf{g}_M)$. 
Since $M$ is conformally flat, there exists a coordinate system
$(X^i)$ such that
$$
\mathbf{g}_M=Q_{ME}^{-1}\sum_i dX^i\otimes dX^i,
$$
where $Q_{ME}$ is the conformal factor transforming $\mathbf{g}_M$ into the flat Euclidean metric $\mathbf{g}_E$.
Then, if we formally replace $(x^1,x^2,x^3)$  by $(X^1,X^2,X^3)$ in the transformations (\ref{Kalnins_formula_b})
we obtain the coordinate transformations from $(X^i)$ to the $R$-separable coordinates $(q^i)$.
Indeed, by inserting these relations in the metric $\mathbf{g}_M$, we have
$$
\mathbf{g}_M=Q_{ME}^{-1}\sum_i dX^i\odot dX^i=Q_{ME}^{-1} Q_E^{-1}\sum_i[P(q^i)\cdot(q^{i+2}-q^{i+1}) dq^i\odot dq^i].
$$ 
Hence, $Q_M=Q_{ME}Q_E$ is the conformal factor that transforms the general conformally flat metric (\ref{m_Rsep}) into a metric on the specific conformally flat manifold $M$.
Then, in order to compute the conformal factor and the coordinate transformation, we only need to know the coordinates $X^i$ on $M$ corresponding to the Cartesian coordinates on $\mathbb E_3$.

In the following example we develop explicitly the case of $\mathbb S_3$.

\begin{nexmp}\label{example_20} \rm
Let $(X^1,X^2,X^3)$ be stereographic coordinates  on $\mathbb S_3$, considered as a submanifold of $\mathbb E_4$. They are related to the Cartesian coordinates
$(x^1,\ldots,x^4)$ of $\mathbb E_4$ by the following equations
$$
\begin{array}{l}
x^a=\dfrac{2r^2X^a}{r^2+ \sum_{i=1}^3 (X^i)^2} \qquad a=1,\ldots,3\\
x^4=r - \dfrac{2r^3}{r^2+ \sum_{i=1}^3 (X^i)^2}
\end{array}
$$
where $r$ is the radius of the sphere. The components of the metric of $\mathbb S_3$ in the coordinates $(X^i)$
are (see also \cite{Eisen_book})
$$
g_{ii}= \frac{4r^4}{(r^2+\sum_{i=1}^3 (X^i)^2)^2}.
$$
Hence, the function $Q_{SE}=(r^2+\sum_{i=1}^3 (X^i)^2)^2/4r^4$ is the conformal factor relating $\mathbb S_3$ 
to $\mathbb E_3$. Then,
$$
Q_S= \frac{\lambda^2[(r^2+\sum_{i=1}^3 (X^i(q^1,q^2,q^3))^2)^2]}{r^4\prod_h(q^h-q^{h+1})}
$$
is the conformal factor which makes (\ref{m_Rsep}) the metric of $\mathbb S_3$. The coordinates $(q^1,q^2,q^3),$ related to the stereographic coordinates $(X^1,X^2,X^3)$ by
$$
\begin{array}{rcl}
\lambda\cdot{X^1} &=& \sqrt{\frac{(q^1 - e_2)\cdot(q^2 - e_2)\cdot(q^3 - e_2)}{(e_2 - e_1)\cdot(e_2 - e_3)\cdot(e_2 - e_4)\cdot(e_2 - e_5)}},\\
\lambda\cdot{X^2} &=& \sqrt{\frac{(q^1 - e_3)\cdot(q^2 - e_3)\cdot(q^3 - e_3)}{(e_3 - e_1)\cdot(e_3 - e_2)\cdot(e_3 - e_4)\cdot(e_3 - e_5)}},\\
\lambda\cdot{X^3} &=& \sqrt{\frac{(q^1 - e_4)\cdot(q^2 - e_4)\cdot(q^3 - e_4)}{(e_4 - e_1)\cdot(e_4 - e_2)\cdot(e_4 - e_3)\cdot(e_4 - e_5)}},
\end{array}
$$
with $\lambda$ given by (\ref{Kalnins_formula_a}), are coordinates on $\mathbb S_3$ in which $R$-separation of the equation
$$
\Delta \psi =\frac 3{4r^2} \psi.
$$
occurs for the fixed value of the energy determined by the radius of the sphere.
\end{nexmp}

\section{Conclusions}\label{sec_5}
The examples show the effectiveness of the concept of fixed energy $R$-separation also to cases
with non-zero energy and justifies the study of the CI-Laplacian. 
A direction for further research lies in the cases of metrics with one or more conformal symmetries, by using the techniques discussed.
Another interesting open problem is the discussion for a general Riemannian manifold of the geometric interpretation of the compatibility condition \ref{Xicon} 
from Theorem \ref{t_Rsep}.

\section*{Acknowledgements}
The authors wish to thank their reciprocal departments for hospitality during periods when parts of this paper were written.  They also wish to express their
appreciation for helpful discussions with Giovanni Rastelli.  The research was supported in part by a Natural Sciences and Engineering Rearch Council
of Canada Discovery Grant (RGM), C.M Lerici foundation, Sweden (MC) and by a Ministero dell'Universit\`a e della Ricerca PRIN project (CMC).

\end{document}